\documentclass[12pt]{iopart}
\input epsf
\usepackage{graphicx} 
  \expandafter\let\csname equation*\endcsname\relax
  \expandafter\let\csname endequation*\endcsname\relax
\usepackage{amsmath} 
\usepackage{amssymb} 
\usepackage{euscript}
\usepackage{mathrsfs}
\newcommand{\pd}{\partial}
\begin{document}

\title{Wormholes and Naked Singularities in Brans-Dicke cosmology}

\author{D.A. Tretyakova}
\address{Department of Celestial Mechanics, Astrometry and Geodesy, Physics Faculty, Lomonosov Moscow State University, Universitetsky pr. 13, Moscow, 119991, Russia}
\address{Department of Theoretical Physics, Physics Department, Institute for Natural Sciences, Ural Federal University, Lenin av. 51, Ekaterinburg, 620083, Russia}

\author{B.N. Latosh}
\address{Faculty of Natural and Engineering Science, Dubna International University, Universitetskaya Str., 19, Dubna, Moscow Region, Russia}
\address{Bogoliubov Laboratory of Theoretical Physics, Joint Institute for Nuclear Research, Joliot-Curie 6, 141980 Dubna, Moscow region, Russia}

\author{S.O. Alexeyev}
\address{Sternberg Astronomical Institute, Lomonosov Moscow State University, Universitetsky pr. 13, Moscow, 119991, Russia}

\date{\today}

\begin{abstract}
We perform analytical and numerical study of static spherically symmetric solutions in the context of Brans-Dicke-like cosmological model by Elizalde et al. \cite{Elizalde:2004mq} with an exponential potential. In this model the phantom regime arises without the appearance of any ghost degree of freedom due to the specific form of coupling. For the certain parameter ranges the model contains a regular solution which we interpret as a wormhole in an otherwise dS Universe. We put several bounds on the parameter values: $\omega<0 ,\,\, \alpha^2/|\omega|<10^{-5}, 22.7\lesssim\!\phi_0\!\lesssim25\,$. The numerical solution could mimic the Schwarzschild one, so the original model is consistent with astrophysical and cosmological observational data. However differences between our solution and the Schwarzschild one can be quite large, so black hole candidate observations could probably place further limits on the $\phi_0$ value. 
\end{abstract}


\section{Introduction}
\label{s0}

The astrophysical data (especially from the last decades) ranging from high redshift surveys of supernovae to WMAP observations  indicate that  our Universe is experiencing an accelerating phase of expansion now \cite{2013ApJS19H}. A possible interpretation of this expansion in terms of General Relativity states that about $70\%$ of the total energy of our Universe is attributed to the dark energy  with large and negative pressure (cosmological constant is the best fit nowadays, see  however \cite{Bamba:2012cp} for a comprehensive review on dark energy ore \cite{2013FrPhyL} for a brief one). One of the simplest possibilities of a theoretical description of the dark energy, beyond the cosmological constant, is to add a scalar field into the GR action (a scalar-tensor theory). 

Brans-Dicke (BD) scalar-tensor theory was one of the first attempts to modify gravity \cite{BD1961}. 
This theory introduces a scalar field $\phi$ which manifests itself as gravitational constant variation. The action of BD theory has the form \cite{BD1961}:
\begin{equation}
S=\cfrac{1}{16\pi} \int d^4 x ~ \sqrt{-g} \left[  \phi \mathcal{R} - \cfrac{\omega}{ \phi} \partial_\mu \phi \partial^\mu \phi  \right],
\end{equation}
Through the years the theory passed multiple observational and theoretical tests.  A variety of physical phenomena  can be interpreted in terms of Brans-Dicke theory, ranging from the flat galaxy rotation curves \cite{Riazi:1993mt}  to inflation \cite{Herrera:1995me} and cosmological constant problem (vacuum catastrophe) \cite{Klebanov:1988eh}. It is now also well established that Brans-Dicke  theory can be used for dark energy  modelling \cite{PhysRevD.63.043504,de1,Hrycyna:2013hla}. 

The current observational bound on the BD parameter $|\omega|>50000$ is imposed in terms of the parameterized post-Newtonian (PPN) expansion of the theory \cite{Bertotti2003}. Such a large value causes debates on the theory's adequacy since it's PPN expansion reaches general relativity (GR) in the limit $\omega\to\infty$ (see \cite{Faraoni1999yp} for detailed information on connection between BD and GR). Nevertheless, papers \cite{PhysRevLett.70.2217}-\nocite{PhysRevD.48.3436, PhysRevD.59.123502}\cite{PhysRevD.66.046007} show that most of the scalar-tensor theories in their cosmological evolution reach the state where the contribution of the scalar part to gravity virtually vanishes. So although $\omega$ is large nowadays, it could demonstrate more interesting values (from the observational point of view) in the past. On the other hand, regardless of the value of $\omega$ (paper \cite{Tretyakova2011ch} proposes $|\omega|>10^{40}$) for $|\omega|<\infty$ the theory can describe bouncing cosmology: considering the evolution of the Universe backwards in time, one can see that the scale factor does not turn to zero at any moment.  Its value decreases to a minimum (bounce) and then starts increasing again. All the characteristic functions remain regular at the bounce \cite{Tretyakova2011ch}, contrary to the presence of the initial singularity in the classical GR cosmology. Hence the theory is of scientific interest since large $\omega$ simply provides the agreement with observational data and also helps avoiding the cosmological initial singularity.

In the BD theory the possible way to take dark energy into account is to add a scalar field potential to the action \cite{Kamenshchik:2012pw,Pozdeeva:2014isa}. As it was shown in \cite{Hrycyna:2013hla} for an arbitrary form of the scalar field potential the Universe evolution could mimic the $\Lambda$CDM model (i.e. observational data) in BD theory. In this paper we consider spherically symmetric solutions in the BD-like model arising in the cosmological context in the paper by Elizalde et al. \cite{Elizalde:2004mq}. The action has the form: 
\begin{eqnarray}
S=\cfrac{1}{16\pi} \int d^4 x ~ \sqrt{-g} e^{\alpha \phi} \left[ \mathcal{R} - \omega \partial_\mu \phi \partial^\mu \phi - V_0 e^{\phi/\phi_0}  \right]. \label{action}
\end{eqnarray}
Here $\alpha$ and $\phi_0$ are constants. 
In such a model the phantom regime arises in a BD-type scenario without the appearance of any ghost degree of freedom due to the specific form of the coupling. It is also shown in \cite{Elizalde:2004mq} that quantum gravity effects may prevent (or, at least, delay or soften) the otherwise unavoidable finite-time future singularity (Big Rip), associated with the phantom. Although this action can seem quite ad-hoc, it can naturally emerge in the low-energy limit of string/M-theory \cite{Foffa:1999dv, Elizalde:2004mq}, which means that the cosmological phantom can emerge naturally within the multidimensional unification theories. All this makes the consideration of this model very attractive from the theoretical point of view. 

Elizalde et al. consider late-time cosmology with equation of state parameter value close to $-1$ (this value is favored by observations \cite{2013ApJS19H}). Under these circumstances an exact spatially-flat Friedman-Robertson-Walker (FRW) cosmology is constructed admitting acceleration phases for the current universe. The model agrees with the observational data in a wide parameters range. Such behaviour is not spoiled at the perturbation level since the scenario is free of perturbative instabilities \cite{Elizalde:2004mq}.  All the above arguments make this model perspective for further versatile research.

Negative values of $\omega$ are often ruled out because of the wrong sign before the kinetic term in the action leading to a ghost. Nevertheless a plenty of researchers consider $\omega<0$ for the following reasons. The latest observational data indicate that phantom nature of dark energy is more likely \cite{2013ApJS19H}. Combining WMAP+eCMB+BAO+H0+SNe data yields to $w_{DE_0} = -1.17(+0.13 -0.12)$ for the flat Universe at the significance level of $68\%$. Such a trend also occurs for a non-flat Universe model and for different dataset combinations. Finally, consideration of a scalar field as an effective description for the theory with positive defined energy could eliminate the quantum contradictions \cite{Nojiri:2003vn, Carroll2003st}. All these arguments make the consideration of BD theory with $\omega<0$ reasonable, since it leads to a phantom cosmology. Modern observational bounds limit the absolute value of $\omega$. Modelling the gravitational collapse does not rule out negative $\omega$ values as well \cite{ PhysRevD.51.4236}. So the range $\omega<0$ is of great interest from the phenomenological point of view, since it can provide no-ghost but phantom cosmology in agreement with modern observations.

In this paper we explore static spherically symmetric solutions of (\ref{action}). Cosmology studies scalar field time dependence $\phi=\phi(t)$, this approximation stands for cosmological homogeneity. Understanding the local effects $\phi=\phi(r)$ yields to a picture of field behaviour at different scales and checks for model consistency. Exploring the cosmological model from another point of view is a step towards a better understanding of dark energy.

The most well-known static spherically symmetric solution in BD was  obtained by Campanelli and Lousto \cite{Campanelli1993sm}. This solution was independently rediscovered by Agnese and La Camera \cite{Agnese2001ci} and correctly reinterpreted as a naked singularity for $\gamma<1$ and a wormhole for $\gamma>1$ (here $\gamma$ is a PPN parameter). Scalar field plays the role of the exotic matter at the wormhole throat and ensures it's traversability not only for $\omega<0$ but also for large positive $\omega$ \cite{Nandi1997mx}. Astrophysical properties were studied by Alexeyev et al. in \cite{alekseev2011R} and found to be in agreement with modern observations.

The purpose of this paper is to explore the properties of the  static spherically symmetric solutions in the framework of Elizalde et al. To do this we analytically define the parametrization for the pure Agnese and La Camera solution and obtain a new analytical solution for the specific metric ansatz, which can be called a ``stealth Schwarzschild solution'' (analogously to \cite{Babichev:2013cya}).  The presence of the scalar field potential modifies the field equations, so new types of solutions can emerge. We perform a numerical exploration of this solutions and discuss their properties. Elizalde et al. construct a model in such a way that the ghosts can be absent, so it is interesting to check whether regular local solutions can appear in the ghost-free sector of the model.

Spherically symmetric solutions for similar actions are present in the literature. An asymptotically Lifshitz black hole solution for a power-law potential in BD setting was found in \cite{Maeda:2011jj} but the allowed $\omega$ parameter range appears to be too narrow to fit the observations. Another wormhole solution for the potential of the form $V(\phi)=\Lambda\phi$ is given in \cite{Xiao:1991nv}. The latter one is obtained for $\phi=\phi(t)$, this wormhole is non-static, its throat radius will increase as the cosmic time increases.  For the best of our knowledge these are the only solutions in the BD setting with a scalar filed potential. The paper \cite{Garay:1992ej} considers quantum wormhole solutions for the BD gravity with $\Lambda$ in the absence of matter perturbatively, from the field theory point of view. Although  the  framework of \cite{Garay:1992ej} is similar to our and can be obtained via $V(\phi)=2\Lambda$, the results of \cite{Garay:1992ej} cannot be applied to astrophysics. Wormhole in solutions the GR setting with exotic matter and $\Lambda$ representing similar behaviour were found in \cite{Lemos:2003jb,Heydarzade:2014ada}.  The cosmological constant affects the wormhole and its properties, the wormhole geometry will be dominated by the de Sitter space far from the throat. Our  solution is new and differs from the ones of \cite{Lemos:2003jb,Heydarzade:2014ada} since the role of the exotic matter is played by the scalar field.
 
The outline of our paper is the following. In Section \ref{s2} we study the generic properties of discussed BD-like solutions, Section \ref{s3} is devoted to specific analytical solutions, in Section \ref{s4} we obtain some new limitations on model parameters, Section \ref{s5} is devoted to ISL modifications, in Section \ref{s6} we show the results of numerical study. Section \ref{s6} contains discussion and conclusions.  

\section{General properties}
\label{s2}
The corresponding Einstein and Klein-Gordon equations for the action (\ref{action}) have the following form:
\begin{eqnarray}
&&\!\!\!\!\!\!\!\!\!\!\!\!\!\!\!\!\!\!\!\!\!\!\!\!
\alpha \mathcal{R} + \omega \alpha \partial_\mu \phi \partial^\mu\phi - V_0 e^{\phi/\phi_0}\left(\alpha+\cfrac{1}{\phi_0}\right) +  2\omega \square \phi =0 ~,\label{kg0}\\
&&\!\!\!\!\!\!\!\!\!\!\!\!\!\!\!\!\!\!\!\!\!\!\!\!
G_{\mu\nu}=\omega \left(\partial_\mu\phi \partial_\nu \phi -\cfrac{1}{2}g_{\mu\nu} \partial_a\phi \partial^a\phi\right)+e^{-\alpha\phi}\left(\nabla_\mu\nabla_\nu e^{\alpha\phi}-g_{\mu\nu}\square e^{\alpha\phi}\right)-\cfrac{1}{2}g_{\mu\nu}V(\phi). \label{jor}
\end{eqnarray}

Hawking's theorem on BD theory \cite{hawking1972} is usually taken to mean that BD static spherically symmetric solutions are exactly the same as those of GR. We want to note that this is an overstatement. For example, the black hole solution of Campanelli and Lousto \cite{Campanelli1993sm} is spherically symmetric and static but does not match the Schwarzschild one. The proof of Hawking's theorem relies on the equivalence between Jordan and Einstein frames and the fact that the corresponding conformal transformation is well-defined. As it is also pointed out in \cite{hawking1972} the advantage of using the Einstein frame is that in the Einstein frame BD scalar has canonical kinetic energy density and obeys the weak (WEC) and null (NEC) energy conditions. If the conformal transformation to the Einstein frame becomes ill-defined at the horizon (as it occurs in \cite{Campanelli1993sm}) then its variables ${g_{E\mu\nu},\phi_E}$ (subscript $E$ stands for Einstein frame) cannot be used on such surface \cite{Faraoni2010rt}. The last statement invalidates the proof of Hawking's theorem. One could use, of course, the Jordan frame instead of the Einstein one. In this case the scalar $\phi$ violates the WEC  and NEC because its stress-energy tensor has a non-canonical structure containing second derivatives of $\phi$ instead of being quadratic in the first derivatives.

So, if our Einstein frame scalar does not obey the WEC or the Jordan frame scalar diverges at the horizon then the Hawking theorem does not apply and the solution is not forced to be Schwarzschild one. 

One can perform the transformation to the Einstein frame $g_{\mu\nu}e^{\alpha\phi}=g_{\mu\nu}\Omega^2=g_{E\mu\nu}$.
\begin{eqnarray}
&&S_E=\cfrac{1}{16\pi} \int d^4 x ~ \sqrt{-g_E} \left[ \mathcal{R}_E - \tilde{\omega} \partial_\mu \phi \partial^\mu \phi - \tilde{V}(\phi)\right]~,\\
&&ds^2_E = e^{\nu_E} dt_E^2 - e^{\lambda_E}dr^2_E - r^2_E (d\theta^2 + \sin^2\theta d\varphi^2) ~,\\
&&\tilde{\omega} =(3\alpha^2/2+\omega), \qquad \tilde{V}(\phi)=Ve^{-\alpha\phi} =V_{E0} e^{-\phi/\phi_{E0}}. \label{ef}
\end{eqnarray}
It follows from the expression for $\tilde{\omega}$ (\ref{ef}) that even if $\omega$ is negative, in the case when 
\begin{equation}
3\alpha^2/2+\omega >0 \label{noghost}
\end{equation}
the effective kinetic energy of $\phi$ becomes positive, similarly to the usual scalar field, and therefore the ghost does not appear, although the cosmology is phantom \cite{Elizalde:2004mq}.
Introducing the four-velocity $\tilde{u}^a=\left(e^{-\nu_E/2},0,0,0\right)$ we obtain the WEC in the form
\begin{equation}
\tilde{\omega}\phi'^2/\phi^2 e^{-\lambda_E}-\tilde{V}(\phi)\geq 0, \label{wec}
\end{equation}
which is generally satisfied  everywhere if $V_{E0}\leq 0$ and $3\alpha^2+2\omega>0$.

Via redefinition $\Phi=e^{\alpha\phi}$ one can obtain the theory in the form of the standard BD one plus potential:
\begin{eqnarray}
S=\cfrac{1}{16\pi} \int d^4 x ~ \sqrt{-g} \left[ \Phi \mathcal{R} - \cfrac{\omega}{\alpha^2} \cfrac{\partial_\mu \Phi \partial^\mu \Phi}{\Phi} - V_0\Phi^{(1+\frac{1}{\alpha\phi_0})} \right]~. \label{redef}
\end{eqnarray}

Having in mind that we are now working with the ``usual'' BD  we can directly apply it's well known weak field limit results to our model. We obtain that $G \sim \Phi^{-1}=e^{-\alpha\phi}$, where $G$ is the effective gravitational constant. 

\section{Analytical solutions}
\label{s3}
Let us rewrite the field equations as in \cite{Campanelli1993sm}. The Klein-Gordon equation \eqref{kg0} can be rewritten as
\begin{eqnarray}
\alpha  R - \omega \alpha \pd_\mu \phi \pd^\mu\phi - V_0 e^{\phi/\phi_0}\left(\alpha+\cfrac{1}{\phi_0}\right) +  \cfrac{2\omega}{\alpha} e^{-\alpha \phi} \square e^{\alpha \phi} =0 ~. \label{kg1}
\end{eqnarray}
Multiplying (\ref{jor}) by $g^{\mu\nu}$  we obtain
\begin{eqnarray}
 R - \omega \pd_\mu \phi \pd^\mu\phi - 2V_0 e^{\phi/\phi_0}  + 3 e^{-\alpha \phi} \square e^{\alpha \phi}=0 ~. \label{kg3}
\end{eqnarray}
Combining  equations \eqref{kg3} and (\ref{kg1}) we get
\begin{eqnarray}
-\left(\cfrac{2\omega}{\alpha}+3\alpha\right) \square e^{\alpha \phi}-V_0 e^{\phi/\phi_0}\left(\alpha-\cfrac{1}{\phi_0}\right)e^{\alpha \phi} =0 ~. \label{kg4_0}
\end{eqnarray}
Substituting (\ref{kg3}) and (\ref{kg4_0}) into the field equation (\ref{jor}) we arrive at the following equation
\begin{eqnarray}
&&R_{\mu\nu}=\omega \partial_\mu\phi \partial_\nu \phi +e^{-\alpha\phi}\nabla_\mu\nabla_\nu e^{\alpha\phi}+
\cfrac{1}{2} g_{\mu\nu}De^{\phi/\phi_0},  \label{app} \\
&& D=V_0\left[ 1-\left(\alpha-\cfrac{1}{\phi_0}\right)\cfrac{\alpha}{2\omega+3\alpha^2}\right]. \label{kg4}
\end{eqnarray}
Further we obtain specific analytic solutions for fixed parameter relations or specific metric structure. Of course, the general analytical solution remains unknown. In section \ref{s6} we investigate the general solution numerically.

\subsection{The case $D=0$}

When $D=0$, or, equivalently
\begin{equation}
\phi_0 =-\cfrac{\alpha}{2(\alpha^2+\omega)}, \label{phi0}
\end{equation}
one has the well-known solution of Agnese and La Camera (ALC) \cite{Agnese2001ci}, since the field equations match in this case. Within the framework of Elizalse et al. this solution looks like
\begin{eqnarray}
&&ds^2 = A(r)^m dt^2 -A(r)^{-n}dr^2 - A(r)^{1-n}r^2 (d\theta^2 + \sin^2\theta d\varphi^2), \label{pwlg}\\
&&\phi =\cfrac{n-m}{2\alpha^2}\ln\left[ F\left(1-\cfrac{2\eta}{r}\right)\right], \quad A(r)=1-\cfrac{2\eta}{r} \label{AC1}\\
&&\eta=M\sqrt{(1+\gamma)/2}, \qquad m=\sqrt{2/(1+\gamma)}, \qquad n=\gamma\sqrt{2/(1+\gamma)},\\
&&\cfrac{\omega}{\alpha^2}=\cfrac{m-2n}{n-m},  \qquad G=e^{-\alpha\phi}\cfrac{2}{1+\gamma}\, , \label{pwlg-2}
\end{eqnarray}
with $\gamma$ being the PPN parameter, $\eta$ plays a role of the mass parameter, $G$  is the effective gravitational constant and  $F$ can be fixed from the Newtonian limit. The parameter $\alpha$ comes from the coupling of Elizalde et al. action \eqref{action}.  This solution corresponds to a naked singularity if $\gamma<1$, to a Schwarzschild black hole for $\gamma=1$ and to a wormhole if $\gamma>1$. The scalar field diverges at the horizon and this is the reason of the Hawking theorem violation. The factor $\sqrt{(1+\gamma)/2}$ is usually absorbed in the mass definition, which results in the correction of order of $10^{-6}$ for $|\gamma-1|\sim 10^{-5}$.

It is worth noticing that the condition $D=0$ preserves the potential term in the action but effectively removes it from the Einstein equations. If we suppose that the scalar field potential acts as the cosmological constant, we can arrive at a conclusion that the accelerated expansion of the Universe does not contribute to local dynamics. This effect is indifferent to the value of the cosmological constant.

To estimate the constant $F$ we must set up the Newtonian limit, in other words, get the model linearised. This work is shown in the Appendix. Comparing (\ref{g0}) with (\ref{AC1}) for large distances we arrive at
\begin{eqnarray}
&&\phi_{\infty}=\cfrac{1}{\alpha}\ln \left(\cfrac{1}{G_0}\cfrac{2\omega+4\alpha^2}{2\omega+3\alpha^2}\right), 
 \qquad F= \left(\cfrac{1}{G_0}\cfrac{2\omega+4\alpha^2}{2\omega+3\alpha^2}\right)^{\cfrac{2\alpha}{n-m}} \label{inf4}
\end{eqnarray}
For the power-law generalization of Schwarzschild (\ref{pwlg}) the equations (\ref{app}) can only be solved for $m=n$ which leads to $D=0$. So there are no other power-law solutions for the ansatz (\ref{pwlg}) except ALC here.

\subsection{Schwarzschild-like solution}

The other solution can be obtained if we suppose that  we measure exactly Schwarzschild metric
\begin{eqnarray}
&&ds^2 = e^{\nu(r)} dt^2 - e^{\lambda(r)}dr^2 - r^2 (d\theta^2 + \sin^2\theta d\varphi^2) ~,\\
&& \nu= \ln(1-r_h/r) \qquad \qquad \lambda=- \ln(1-r_h/r).
\end{eqnarray}
This leads to the solution
\begin{equation}
\phi=\cfrac{\alpha}{\alpha^2+\omega^2}\ln|r+C_1| + C_2, r\neq - C_1 \label{inf3}
\end{equation}
Having in mind the redefinition (\ref{redef}) we see that $G$ is a decreasing function of $r$
\begin{equation}
G \sim (r+C_1)^{-\frac{1}{1+\omega^2/\alpha^2}}. \label{inf5}
\end{equation}
This solution can be called a ``stealth Schwarzschild solution'', analogously to the solution of \cite{Babichev:2013cya} (obtained within the Horndeski/Galileon framework) representing a Schwarzshild metric with non-trivial scalar field. 

To estimate the constants on (\ref{inf5}) we must use the Newtonian limit, in other words, get the model linearised. The scalar $\phi$ increases indefinitely upon $r$, so the theory can only be formally linearised as $\phi=\phi_c+\epsilon$ in the region outside the object $\left | \frac{r}{C_1}\right |<1$ where the Taylor expansion of logarithm exists. The constant $C_1$ represents then the length scale up to which the solution admits Newtonian behavior. 

The obvious condition preserving (to a certain extent) the observational adequacy of the solution is: $ |\omega| \gg |\alpha|$. It makes the power of $r$ small and ensures that $\varphi$ does not change abruptly space and varies less and less with the distance. This condition can also be agreed with (\ref{ppn1}). The space-time is asymptotically flat, but the gravitational constant vanishes at the infinity.

The discovery of the discussed solution under the assumption that it realizes in nature, seems tricky, since the metric is exactly the Schwarzschild one. It only manifests itself as a gravitational constant lowering itself with distance. One could naively suggest that for astrophysical scales this effect would simply hide itself into mass definition, since accurate mass measurements are done mostly for several binary systems due to their orbital motion \cite{2014JCAP055B}. Therefore a lower but very slowly changing gravitational constant value would just be absorbed into calculated masses: the less would be the distance, the larger the object masses would seem  to us. This situation contradicts to the dark matter effect: if we suppose it to be dynamical and not caused by some real-life matter. 

\section{Parameter estimations}
\label{s4}

From the paper by Elizalde et al.  \cite{Elizalde:2004mq} we can get the cosmologically imposed condition, providing the accelerated expansion of the Universe:
\begin{equation}
\phi_0^2>\cfrac{4}{3\alpha^2+2\omega}.\label{cos1}
\end{equation}
For $|\omega| \gg |\alpha^2|$ and $\omega$ is negative this condition is always satisfied  and is also in agreement with (\ref{phi0}). This means that ALC solution can represent a wormhole embedded in an expanding Universe, but not affected by this expansion at all.

The other possible parameter estimation comes from the PPN expansion. The observational bound is \cite{Bertotti2003} $\gamma-1=(2.1 \pm 2.3) \times 10^{-5}$. Following \cite{Damour:1995kt} we derive the PPN-parameters $\gamma$ and $\beta$:
\begin{equation}
\gamma-1=-\cfrac{\alpha^2}{2\alpha^2+\omega}, \qquad \beta-1\equiv 0 \label{ppn1}
\end{equation}
When $|\omega| \gg |\alpha^2|$, taking into account $|\gamma-1| <10^{-5}$ we obtain
\begin{equation}
\alpha^2/|\omega|<10^{-5}. \label{ppn2}
\end{equation}
This requirement disagrees with no-ghost condition (\ref{noghost}) for negative $\omega$. In fact the inequality above leaves no room for the no-ghost sector of the model. This result is a bit disappointing, but is not critical in view of the reasons regarding negative $\omega$ given in the introduction.

\section{Inverse-square law modifications}
\label{s5}

The power-law modification of the inverse-square law (ISL) in the general form reads
\begin{equation}
\mathfrak{V}(r) = \cfrac {G_0 M}{r} \left[1+ a_N \left(\cfrac{\eta}{r}\right)^{N-1}\right].
\end{equation}
Expanding the ALC \eqref{pwlg}-\eqref{pwlg-2} solution $g_{00}$ in binomial series and going to physical units we get
\begin{eqnarray}
&&\mathfrak{V} = \sqrt{\cfrac{2}{1+\gamma}} \frac {\eta c^2}{r} \left[1+\left(1-\sqrt{\cfrac{2}{1+\gamma}}\right)\cfrac{\eta}{r}\right]= \cfrac {G_0 M}{r} \left[1+ a_2 \cfrac{\eta}{r}\right],\\
&&a_2=\left(1-\sqrt{\cfrac{2}{1+\gamma}}\right), \qquad \eta=\sqrt{\cfrac{1+\gamma}{2}} \cfrac{MG_0}{c^2}
\end{eqnarray}
for $N=2$\footnote{The interesting paper \cite{isl4} is worth mentioning here. The authors calculate advances in the perihelia of planets in the solar system by treating the power-law correction as a small disturbance and then connect them with the data of  INPOP10a (IMCCE, France) and EPM2011 (IAA RAS, Russia) ephemerides(by means of minimizing the sum of squared deviations). They take the uncertainty in the Sun's quadrupole moment into account and estimate it along with the parameters of the power-law correction. The result is $N=0.605$ for the exponent. However, from EPM2011, they find that, although it yields $N= 3.001$, the estimated uncertainty in the Sun's quadrupole moment is much larger than the value of $\pm 10\%$ given by current observations. So, no any certain conclusions on the most plausible value of $N$ can be done.}.
We are interested in the case when $\gamma>1$, so $a_2>0$. It is worth noticing that the correction is mass-dependent. Papers \cite{isl1,isl2,isl3} claim the following $1\sigma$ limits for  power-law corrections
\begin{eqnarray}
N=2 \qquad a_N \eta^{N-1} < 1.3\times 10^{-6} [m]. \label{newtb}
\end{eqnarray}
For the conditions of the experiment \cite{isl2} ($M\approx 10$kg) we estimate 
\begin{equation}
a_2\eta= \cfrac{MG_0}{c^2}\left(\sqrt{\cfrac{1+\gamma}{2}}-1\right)=\left(\sqrt{\cfrac{1+\gamma}{2}}-1\right)\times 10^{-16}=\epsilon\times 10^{-16}. \label{epsisl}
\end{equation}
Together with (\ref{newtb}) this gives us $\epsilon<10^{10}$. For the PPN parameter $|\gamma-1|<2.5\times10^{-5}$ we can estimate $\epsilon<2.5\times10^{-6}$ which is a sufficiently stronger bound. This means that one needs to increase the experiment accuracy by 16 orders to approach the PPN-based limit. The satellite motion in the Earth gravitational field or the planet motion around the Sun could represent a more interesting case  whereas $M_{\odot}\sim 10^{30} kg,\quad M_{\oplus}\sim 10^{24} kg$.

Following \cite{proc-disc-2009} we estimate additional relative frequency shift $\left(\delta f/f\right)_p$ for the potential above and make bounds using the frequency measurement accuracy for Galileo Navigation Satellite System (GNSS)\footnote{$h= 23.222\times 10^{3} km, M_{\oplus}=5,972\times 10^{24} kg, R_{\oplus}=6371 km, G_0=6.67384 \times 10^{-11} m^3 kg^{-1} c{-2}$}.
\begin{equation}
\left(\cfrac{\delta f}{f}\right)_p\approx \cfrac{M_{\oplus} G_0}{c^2}a_2\eta\left(\cfrac{1}{ R_{\oplus}^2}-\cfrac{1}{(R_{\oplus}+h)^2}\right)=
\cfrac{M_{\oplus}^2 G_0^2}{c^4}\epsilon\left(\cfrac{1}{ R_{\oplus}^2}-\cfrac{1}{(R_{\oplus}+h)^2}\right)
\end{equation}
Considering the accuracy of the frequency measurement of the Galileo constellation  $\varepsilon_{f_r}=10^{-12}$  analogously to (\ref{epsisl}) we obtain
\begin{equation}
\delta f/f\approx 5 \epsilon\times 10^{-13} <\varepsilon_{f_r}.
\end{equation}
The resulting bound appears to be as $\epsilon < 2$. It is also much weaker than the PPN-based one. This result seems strange, since the experimental technique which provides the PPN and the GNSS bounds is similar. This peculiarity requires additional research.

\section{Numerical results}
\label{s6}

Now we intend to analyse the general solution. For numerical calculations we use the following ansatz
\begin{equation}
ds^2=\Delta dt^2 - \cfrac{\sigma^2}{\Delta} dr^2 - R(r)^2 d\Omega^2 ~. \label{nummet}
\end{equation}
Without loss of generality we set $\sigma=1$ since this is equivalent to the radial coordinate transformation. From the equation (\ref{app}) we obtain the system of differential equations, which can be resolved with respect to highest derivatives as
\begin{eqnarray}
&&\!\!\!\!\!\!\!\!\!\!\!\!\!\!\!\!\!\!\!\!\!\!\!\!\!
\Delta ''(r)=\frac{D R(r) e^{\frac{\phi (r)}{\phi _0}}-2 \Delta '(r) R'(r)}{R(r)} ~,\\
&&\!\!\!\!\!\!\!\!\!\!\!\!\!\!\!\!\!\!\!\!\!\!\!\!\!
R''(r)=\frac{-2 R(r) \Delta '(r) R'(r)-2 \Delta (r) R'(r)^2+D R(r)^2 e^{\frac{\phi (r)}{\phi _0}}+2}{2 \Delta (r) R(r)} ~, \\
&&\!\!\!\!\!\!\!\!\!\!\!\!\!\!\!\!\!\!\!\!\!\!\!\!\!
\phi ''(r)=-\frac{\left(\alpha ^2+\omega \right) \phi '(r)^2}{\alpha }+\frac{2 \left(\Delta (r) R'(r)^2-1\right)}{\alpha  \Delta (r)  R(r)^2} - \frac{D R(r) e^{\frac{\phi (r)}{\phi _0}}-2 \Delta '(r) R'(r)}{\alpha  \Delta (r) R(r)}. 
\end{eqnarray}

The most interesting scales from the phenomenological point of view are the astrophysical ones, such as neutron stars binary systems. Hence we will consider a mass of order of the solar one.

Under the requirement for the potential term in the action to transform to a cosmological constant in the limit $r\to\infty, \phi\to\phi_{\infty}$ we obtain that
\begin{equation}
V_0=2\Lambda e^{-\phi_{\infty}/\phi_{0}}, \qquad \phi_{\infty}= \cfrac{1}{\alpha}\ln\left[ \cfrac{1}{G_0}\cfrac{2\omega+4\alpha^2}{2\omega+3\alpha^2}\right], \label{num2}
\end{equation}
where $\Lambda$ is the cosmological constant. Asymptotic value of the scalar field is derived from (\ref{g0}), $\phi_0$ is a free parameter. The initial values for the numerical integration are taken from the ALC solution. Summarizing the known bounds we can state that $\omega<0$ (no naked singularity in ALC), $\alpha^2 < 10^{-5} |\omega|$ (PPN bound). For the sake of illustration we take  $\omega=-10000$, $\alpha=0.1, M=M_{\odot}$\footnote{$c=2.998\times 10^{8}$ m/sec, $M_\odot=1.988435\times 10^{33}$ g}. 

\begin{figure}[!hbtp]
$$
\includegraphics[keepaspectratio,width=0.6\textwidth]{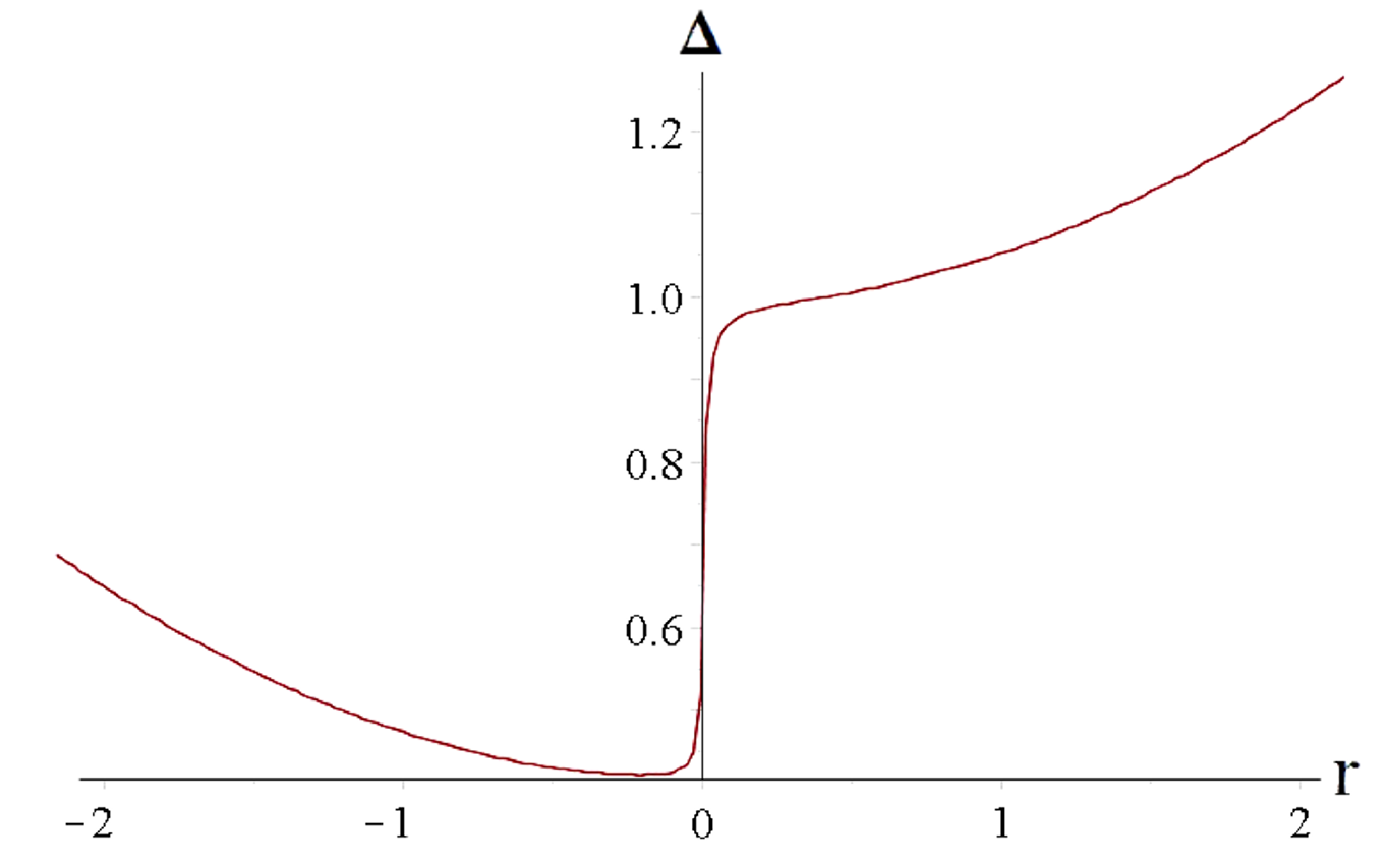}
$$
\caption{The solution represents a wormhole embedded in an otherwise de Sitter universe ($\phi_0=24.5$). Near the throat, the wormhole geometry is dominant while as the distance increases the structure of dS reveals itself.  }
\label{dswh}
\end{figure}

Since the values of $\omega$ and $\alpha$ under consideration provide WEC violation, and requiring the correspondence with Agnese and La Camera solution for small potential influence we claim that the solution represents a wormhole embedded in an otherwise de Sitter universe (see fig. \ref{dswh}).  Near the throat, the wormhole geometry is dominant while as the distance increases the structure of dS reveals itself. 

\begin{figure}[!hbtp]
$$
\includegraphics[keepaspectratio,width=\textwidth]{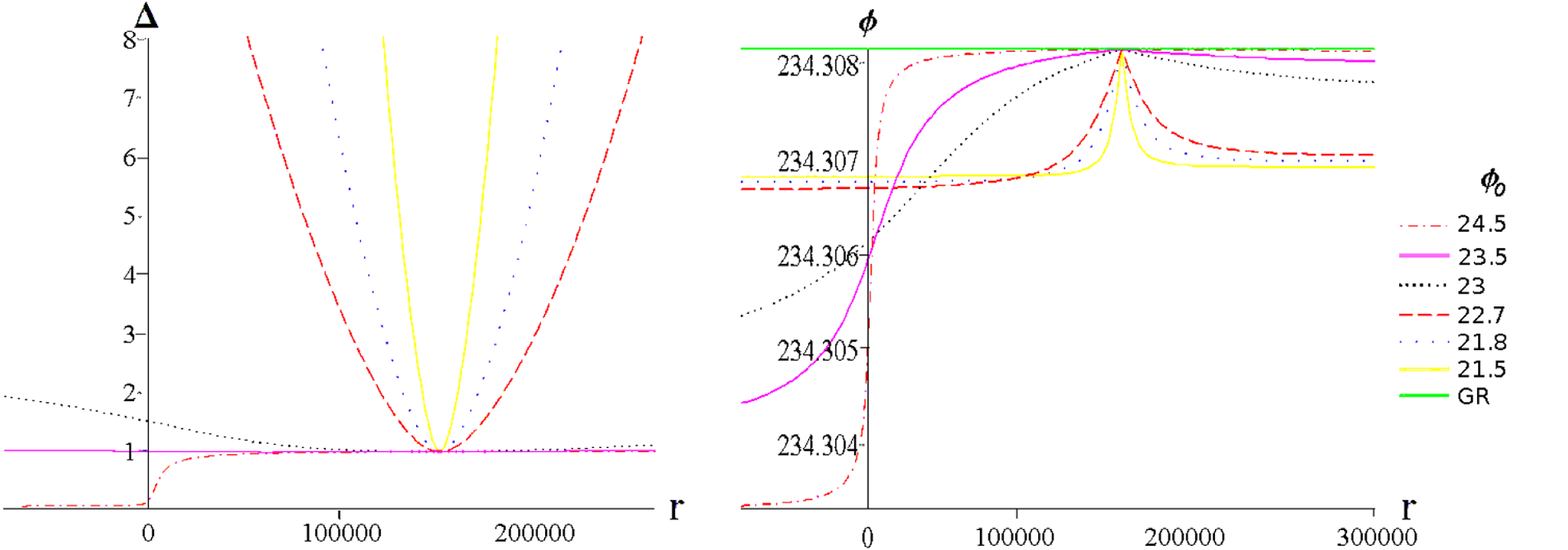}
$$
\caption{Figure represents the enumeration of $\phi_0$ values. Approximately at $\phi_0\approx 22.7$ the solution turns from de Sitter to a wormhole--like. De Sitter solution describes the expanding universe. Wormhole--like solution describes an astrophysical object. Thus the solution can describe both cosmological and astrophysical scales.}
\label{V1_15}
\end{figure}

The space-time approaches the Schwarzschild metric when one continues increasing $\phi_0$ (this corresponds to reducing the potential influence) and differs slightly from Schwarzshild for $\phi_0>25$.  For $25\lesssim\phi_0\lesssim 22.7$ the solution represents a dS-wormhole.  Lowering $\phi_0$ further (strengthen the potential influence)we see that space-time approaches dS. For $\phi_0\gtrsim 22.7$ the wormhole throat does not form, suppressed by the potential term influence (dS case, see fig. \ref{V1_15}). 
De Sitter solution describes the expanding universe. Wormhole--like solution describes an astrophysical object. Thus the solution can describe both cosmological and astrophysical scales.

\begin{figure}[!hbtp]
$$
\includegraphics[keepaspectratio,width=0.8\textwidth]{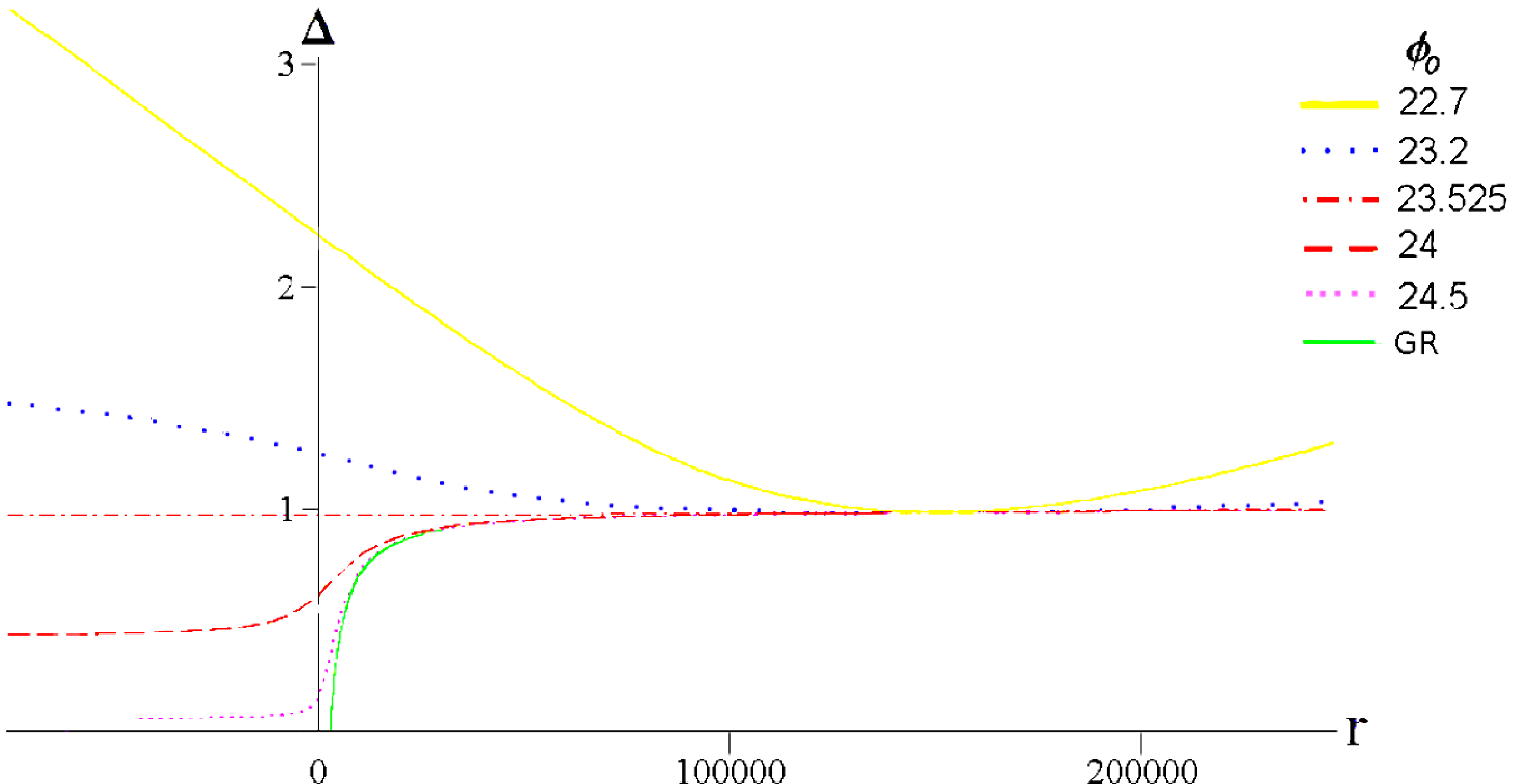}
$$
\caption{Figure represents the enumeration of $\phi_0$ values. The threshold on $\phi_0$ between dS and wormhole-like geometry appears near $\bar{\phi_0}=22.7$.}
\label{VL_5}
\end{figure}

\begin{figure}[!hbtp]
$$
\includegraphics[keepaspectratio,width=\textwidth]{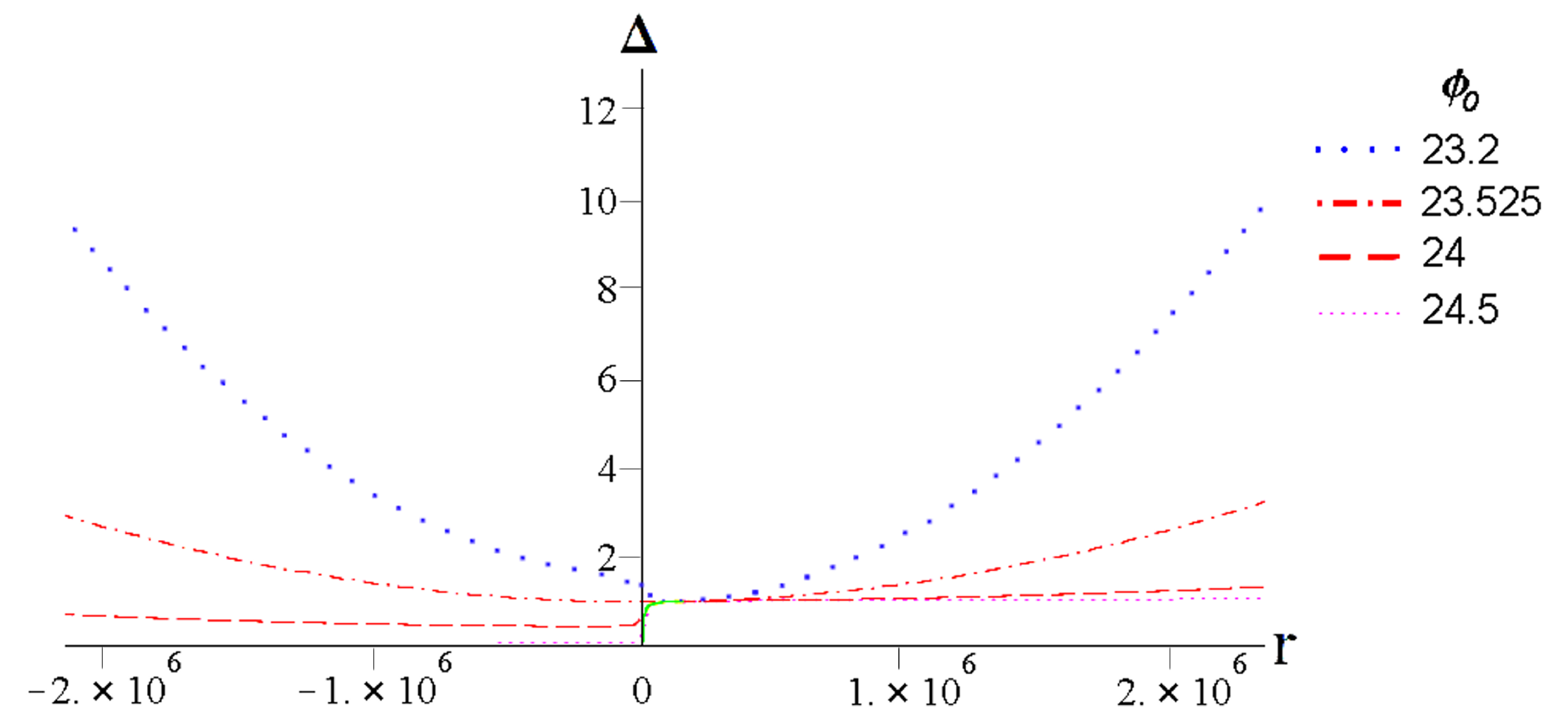}
$$
\caption{The solution represents a wormhole embedded in an otherwise de Sitter universe ($\phi_0=24.5$). Near the throat, the wormhole geometry is dominant while as the distance increases the structure of dS reveals itself.  }
\label{dswh-fahr}
\end{figure}

Figures \ref{VL_5} and \ref{dswh-fahr} display that the given numerical solution is a wormhole analogue of the Schwarzschild-dS solution. Near the throat, the wormhole geometry is dominant (fig. \ref{VL_5}) while as the distance increases the structure of dS reveals itself (fig. \ref{dswh-fahr}).  The threshold on $\phi_0$ between dS and wormhole-like geometry appears near $\bar{\phi_0}=22.7$ (See Fig. \ref{VL_5}).

Solutions representing similar behaviour were found in \cite{Lemos:2003jb,Heydarzade:2014ada}. The papers \cite{Lemos:2003jb,Heydarzade:2014ada} consider the GR setting with exotic matter and $\Lambda$. The cosmological constant affects the wormhole and its properties, the geometry  will be dominated by the wormhole  near the throat and by the the de Sitter space far from the throat, as it is for our solution. Our result are in agreement with \cite{Lemos:2003jb,Heydarzade:2014ada}, however the role of exotic matter is played in our case by the scalar field.

\begin{figure}[!hbtp]
$$
\includegraphics[keepaspectratio, width=0.5\textwidth]{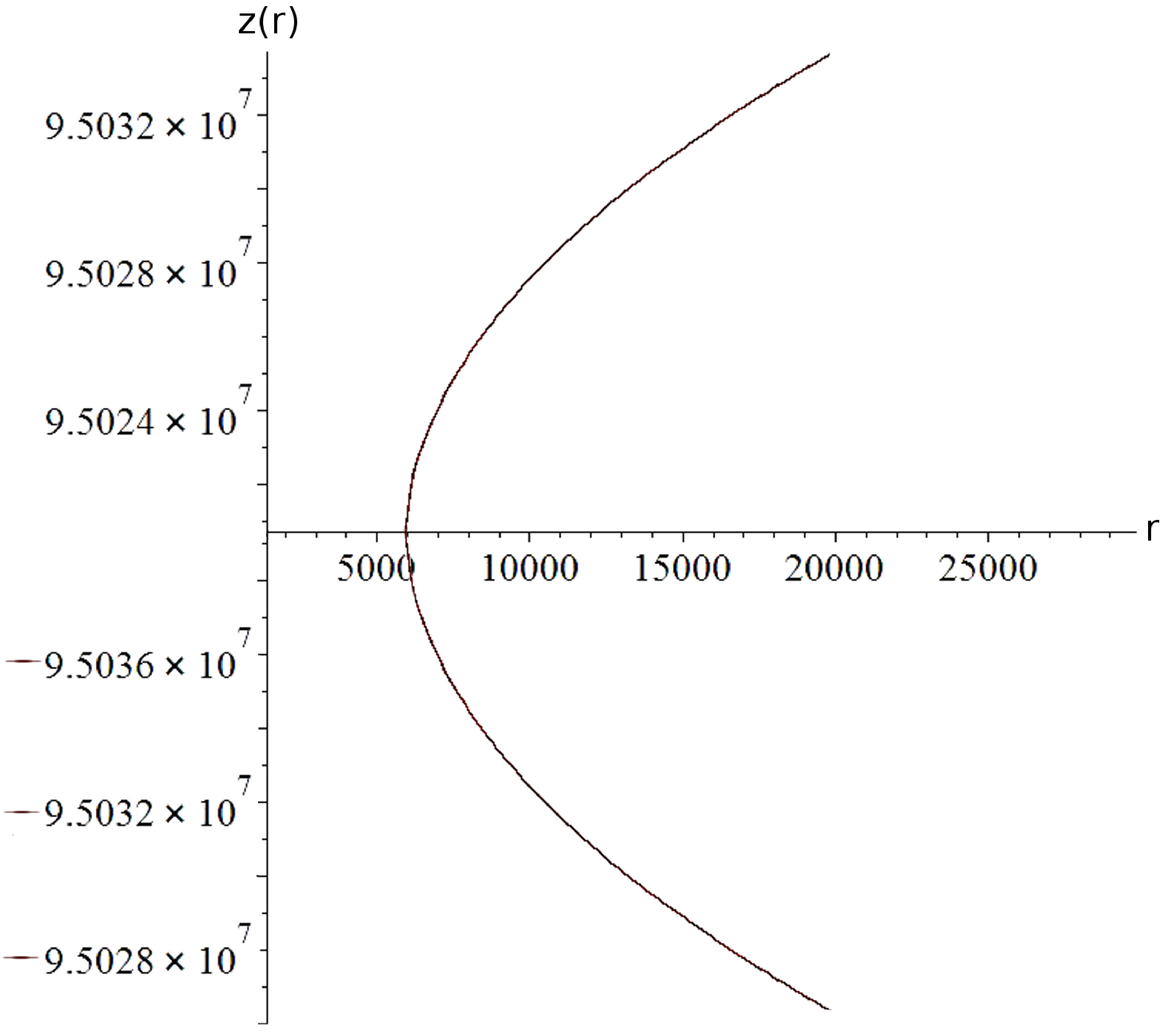}
$$
\caption{The embedded surface is vertical $dz/dr \to \infty $ near the wormhole throat for $\phi_0=24.5$. }
\label{zr}
\end{figure}

Resetting the metric in a form 
\begin{equation}
ds^2=e^{2P(r)} dt^2 - \cfrac{dr^2}{1-b(r)/r} - r^2 d\Omega^2 ~, \label{nummet2}
\end{equation}
we can plot the shape function $b(r)$ and the embedded surface $z(r)$ \cite{Lemos:2003jb}
\begin{equation}
\cfrac{dz}{dr}=\pm\left( \cfrac{r}{b(r)} - 1\right)^{-1/2} ~. \label{emb}
\end{equation}
To be a solution of a wormhole, the geometry has to have a minimum radius, $r =b(r) = r_0$, i.e  the throat, at which the embedded surface is vertical $dz/dr \to \infty $ (see figure \ref{zr}).
For a wormhole to be traversable one must demand the absence of horizons identified as surfaces with $e^{2P}\to 0$, so that the function  $P(r)$ has to be finite everywhere. By the proper choice o $R'(r)$ it is possible to avoid such a divergence at the throat. The numerical calculation shows that for $\phi_0=24.5$ the throat radius is around $r=5927 m$, this is definitely large enough for a microscopical object to pass through. The corresponding ALC throat is at $r=3014 m$, so the potential influence is significant.

The difference between the Schwarzschild solution and numerical solution at the distance around the innermost stable orbit $r=6\eta$  is 
\begin{eqnarray}
&&\Delta_{num}/\Delta_{schw}\approx 1.031 \qquad \phi_0=24.5\,,\label{num3}\\
&&\Delta_{num}/\Delta_{schw}\approx 1.471 \qquad \phi_0=23.5\,.\label{num4}
\end{eqnarray}
These values appear to be quite large, so black hole observations could probably place some limits on $\phi_0$.  There are observational indications that such a deviation might in fact correspond to reality: the paper \cite{Doeleman:2008qh} claims that the apparent size of the source at the galactic centre is less than the expected apparent size of the event horizon of the presumed Sgr A black hole. If this is not due to modelling issues, this could be thought as the indication of GR violation.  To explore the properties of general solution further in detail one requires the analytical form of the solution and we treat this as a task for the future. 

\begin{figure}[!hbtp]
$$
\includegraphics[keepaspectratio, width=0.6\textwidth]{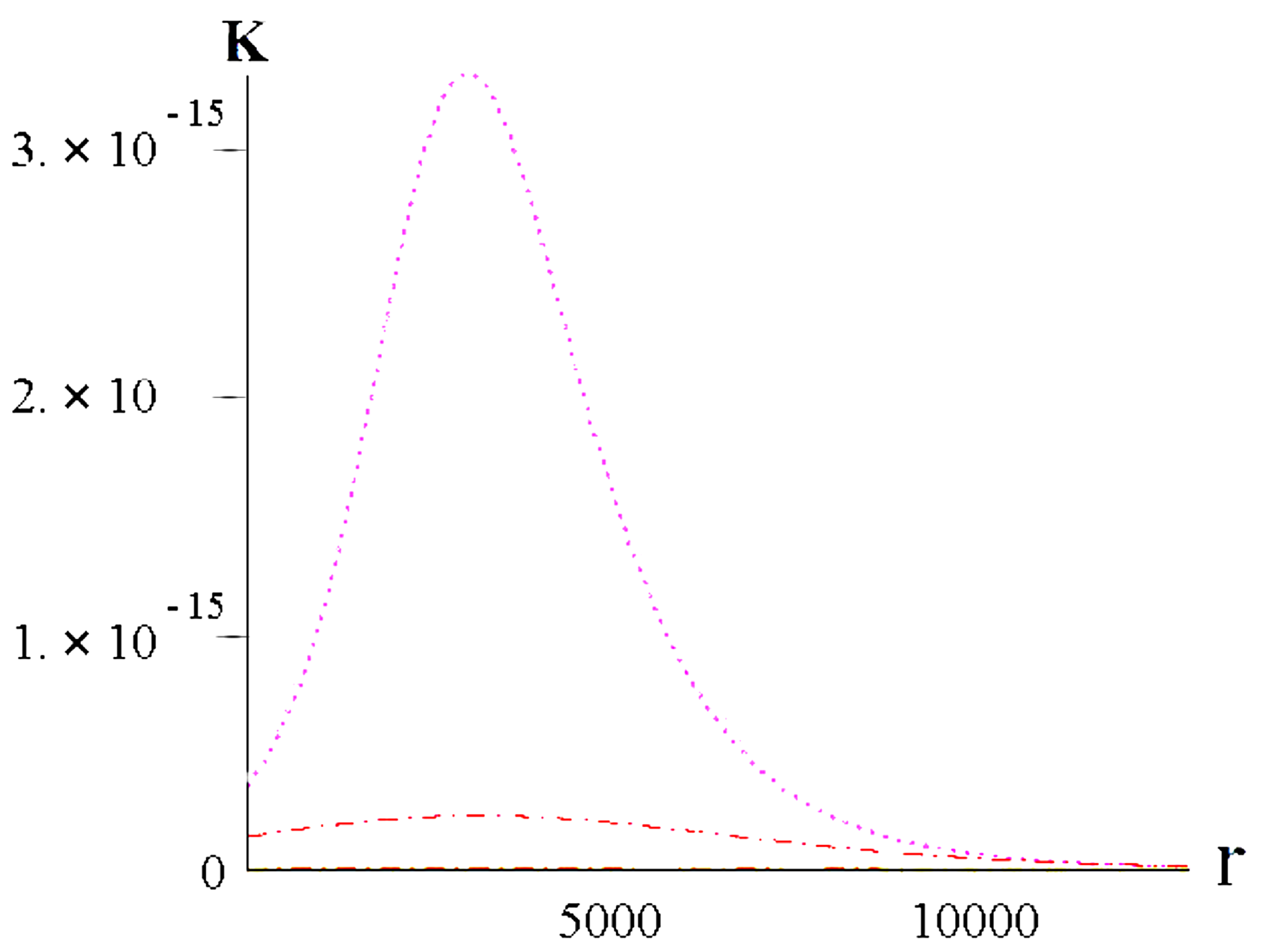}
$$
\caption{The Kretschmann scalar curvature invariant and the metric function $\Delta$ for different parameter values. Only the case $\omega<0 ,\, 22.7\lesssim\phi_0\lesssim 25\,$ represents a regular solution. All other parameter ranges represent a naked singularity for different values of $r$.}
\label{Kretschmann}
\end{figure}

We need to distinguish between the black hole, wormhole and naked singularity cases.  We check different parameter combinations and, compare Kretschmann curvature invariant (see Fig. \ref{Kretschmann}) and metric function $\Delta$ \eqref{nummet}  values. The analysis of multiple plots shows that only the case $\omega<0 ,\, 22.7\lesssim\phi_0\lesssim 25\,$ represents a regular solution: Kretschmann scalar is finite at $r>0$. Therefore this case can represent a traversable wormhole. 

All other parameter ranges seem to represent a naked singularity for different values of $r$ (see Fig. \ref{Kretschmann} also). The cosmic censorship conjecture states  that naked singularities do not emerge in nature.  In terms of wormhole research naked singularities  can be interpreted as  not traversable  ones, so they are of a limited interest. However nowadays the cosmic censorship conjecture remains unproven and many researchers suggest naked singularities as astrophysical object candidates \cite{Kovacs:2010xm, Joshi:2013dva}. In this sense it would be interesting to check for the astrophysical properties of the solution in corresponding parameter regions. We live it for the next step of our investigation.

\section{Discussion and conclusions}
\label{s7}

We explored analytical and numerical properties of static spherically symmetric solutions in the context of BD-like cosmological model. For a certain parameter range $\omega<0 ,\, 22.7\lesssim\phi_0\lesssim 25\,$ the model contains a regular wormhole solution. The space-time geometry is analogously to Schwarzschild-dS one, representing a wormhole in the dS Universe. The influence of the potential term on the wormhole is shown to be be significant. The properties of this newly obtained numerical wormhole solution require further versatile research which is a task for our next paper. 

As it was mentioned in \cite{Elizalde:2008yf} the Jordan and Einstein frames for the theory are not equivalent from the cosmological point of view. It was also argued in \cite{Nandi:1997en} (for BD thery without the scalar field potential) that wormhole solution in the Einstein frame are absent unless  $\tilde{\omega}=3 \alpha^2/2+\omega <0$ (in our notation). As it follows from the equation \eqref{wec} the WEC  can be violated due to the potential alone and one does not need to violate it by setting $\tilde{\omega}<0$ by ``brute force'' \cite{Nandi:1997en}. So, unlike the case of \cite{Nandi:1997en} wormholes are in principle possible in both frames, although the detailed discussion is only possible with the analytical solution in hand.  However the observational bond \eqref{ppn2} implies that the WEC is violated.

From the beginning the discussed model had four free parameters. Original paper \cite{Elizalde:2004mq} provides one useful limitation on $\phi_0$ (\ref{cos1}). We estimate the parameter $V_0$ using cosmological constant and currently measured gravitational one values (\ref{num2}) reducing the number of free parameters by one. We also constraint  $\alpha$ and $\omega$ parameters by (\ref{ppn2}). When considered together equations $\omega<0 ,\, 22.7\lesssim\phi_0\lesssim 25\,$, (\ref{cos1}), (\ref{ppn2}) and (\ref{num2}) put significant constraints on the parameter values in compared with \cite{Elizalde:2004mq}.

We demonstrate that when the equations (\ref{cos1}), (\ref{ppn2}) and (\ref{num2}) are taken into account the model gives rise to a potentially traversable wormhole or a naked singularity. Moreover, since the numerical solutions can mimic the Schwarzschild one, the original model \cite{Elizalde:2004mq} is consistent with the astrophysical observational data. The last statement proofs that the model could be successfully applied both on cosmological and astrophysical scales. However differences between the numerical solution and the Schwarzschild one can be quite large (an example is the factor $1.471$ in (\ref{num3})-(\ref{num4})), so black hole candidate observations could probably place additional limits on $\phi_0$ in a case if the analytical solution would be obtained. The question is whether the proposed parameters can provide deviations visible for the Event Horizon Telescope (for example) and it requires further investigations. The numerical plotting does not allow one to explore all possible parameter combinations, so there ape probably other parameter rangers, where regular solutions exist, and our result represent just an example of such solutions.

This PPN-based requirement (\ref{ppn2}) disagrees with no-ghost condition (\ref{noghost}) for negative $\omega$ and in fact leaves no room for the no-ghost sector of the model. This result is a bit disappointing, but is not critical because of the reasons regarding negative $\omega$ given in the Introduction.
 
We also derive the power-law correction to the gravitational potential for the solution of Agnese \& La Camera \cite{Agnese2001ci}  and check whether it can be observed in the ground-based experiments ore via measuring relative frequency shift for Galileo Navigation Satellite System (GNSS). We conclude that one needs to increase the experiment accuracy to approach the largest possible correction value.

We find out that ALC solution can represent the wormhole embedded in an expanding Universe, but not affected by this expansion.

In summary, we conclude that BD theory being an effective limit of f(R) gravity \cite{Capozziello201079}, describes a wide range of observed phenomena (ranging from cosmological to astrophysical scales), fits the observational data quite well, removes cosmological initial singularity and provides a scale factor bounce in the early Universe. So taking into account all the other results on BD study we would like to state that it is one of the finest candidates for the foundation stone of modern cosmology and extended gravity.
Of course the scalar-tensor theories are not free of problems, especially when they are directly considered as dark energy candidates.  Nevertheless  the attention to phantom models  is  driven by the lack of a good theoretical understanding of the
present universe in terms of  more usual theories. The phenomenology which emerges from these models is rich and 
pertinently at times, so this models deserve to be investigated comprehensively.

\section*{Acknowledgements}
This work was partially supported by individual grants from Dmitry Zimin Foundation ``Dynasty'' (S.A. \& B.L.). The authors are sincerely grateful to A.A. Shatskiy and I. Volkovets for useful discussion. The author are also grateful to the anonymous referees for paying so much attention to the paper and making a huge contribution to it's improvement.

\section*{Appendix I}
\label{s8}
Equation (\ref{app1}) in the presence of matter transforms into
\begin{equation}
R_{\mu\nu}=\omega \partial_\mu\phi \partial_\nu \phi +\cfrac{\nabla_\mu\nabla_\nu e^{\alpha\phi}}{e^{\alpha\phi}}+
\cfrac{1}{2} g_{\mu\nu}De^{\phi/\phi_0} +\cfrac{8\pi}{ e^{\alpha \phi}}\left[ T_{\mu\nu} -\cfrac12 g_{\mu\nu}\left( \cfrac{2\omega+2\alpha^2}{2\omega+3\alpha^2}\right) T^{\mu}_{\mu}\right], \label{app1}\\
\end{equation}
In the weak field limit we can write down
\begin{equation}
\phi=\phi_c+\psi(r), \quad \left |\cfrac{\psi}{\phi_c} \right |\ll 1, \quad e^{ \alpha\phi} \approx e^{ \alpha\phi_c}, \quad \square e^{ \alpha\phi} \approx e^{ \alpha\phi_c}\alpha\square \psi,  \quad g_{\mu\nu}=\eta_{\mu\nu}+h_{\mu\nu}
\end{equation}
where $\eta_{\mu\nu}$ is Minkowski metric and $h_{\mu\nu}$ are perturbation. Up to the first expansion order one obtains:
\begin{eqnarray}
&&R_{\mu\nu}^{(1)}= \alpha\nabla_\mu\nabla_\nu \psi+\eta_{\mu\nu}V_1+ 8\pi e^{-\alpha \phi_c}\left[ T_{\mu\nu} -\cfrac12 \eta_{\mu\nu}\zeta T^{\mu}_{\mu}\right],\label{app2}\\
&&V_1=\cfrac{1}{2}e^{\phi_c/\phi_0}D, \qquad \qquad
\zeta=\left( \cfrac{2\omega+2\alpha^2}{2\omega+3\alpha^2}\right).
\end{eqnarray}
On the other hand
\begin{equation}
2R_{\mu\nu}^{(1)}=\square h_{\mu\nu}+h^{\kappa}_{\kappa,\mu,\nu}-h^{\kappa}_{\mu,\kappa,\nu}-h^{\kappa}_{\nu,\kappa,\mu}
\end{equation}
To simplify \eqref{app2} we impose the following conditions, using the weak field version of harmonic coordinate conditions in a case of the constant scalar field:
\begin{equation}
h^{\kappa}_{\kappa,\mu,\nu}-h^{\kappa}_{\mu,\kappa,\nu}-h^{\kappa}_{\nu,\kappa,\mu}=-2\alpha\nabla_\mu\nabla_\nu \psi
\end{equation}
The equation (\ref{app2}) rewrites as
\begin{equation}
\cfrac12\square h_{\mu\nu}=\eta_{\mu\nu}V_1+ 8\pi e^{-\alpha \phi_c}\left[ T_{\mu\nu} -\cfrac12 \eta_{\mu\nu}\zeta T^{\mu}_{\mu}\right]
\end{equation}

Let us take the space-time and the scalar field generated by a static point-mass M. Therefore the corresponding stress-energy tensor is given by
\begin{equation}
T_{\mu}^{\nu}=diag(M\delta(r),0,0,0), \qquad T^{\mu}_{\mu}=M\delta(r)
\end{equation}
Therefore the metric perturbation reads as:
\begin{eqnarray}
&&\square h_{00}=2V_1+ 16\pi e^{-\alpha \phi_c} \cfrac{ \omega+2\alpha^2}{2\omega+3\alpha^2} M\delta(r)\\
&&h_{00}=-\cfrac{2G_0M}{r}+\cfrac{V_1 r^2}{3}, \\
&&G_0=e^{-\alpha \phi_c}\cfrac{2\omega+4\alpha^2}{2\omega+3\alpha^2}. \label{g0}\\ 
\end{eqnarray}
Finally we obtain the weak field limit for the considering metric in the form
\begin{equation}
g_{00}=1-\cfrac{2G_0M}{r}+\cfrac{V_1 r^2}{3}
\end{equation}

This result agrees with the fact that one can consider the model \cite{Elizalde:2004mq} in the form of the standard BD (\ref{redef}) where $G_0$ plays the role of the effective gravitational constant. Such solution reduces to the usual BD one when $\alpha=1,\varphi=e^{\phi}, V_0=0 $. After comparing it with the de Sitter-Schwarzschild metric we conclude that $V_1$ plays a role of the cosmological constant. 
\\ 
\bibliographystyle{unsrt}
\bibliography{mybib_eng}

\end{document}